 \documentclass[final,5p,times,twocolumn]{elsarticle}

\usepackage{lineno}
 \usepackage{graphics}
\usepackage{amssymb}
\journal{$2^{nd}$ Intern. Conf. Frontiers in
Diagnostic Technologies}

\begin{document}

\begin{frontmatter}



\title{Particle acceleration around SNR shocks}


\author{G. Morlino}
\ead{morlino@arcetri.astro.it}
\address{INAF-Osservatorio Astrofisico di Arcetri, Largo E. Fermi, 5,
50125 Firenze, Italy}

\begin{abstract}
We review the basic features of particle acceleration theory around
collisionless shocks in supernova remnants (SNRs). We show how non linear
effects induced by the back reaction of accelerated particles onto the shock
dynamics are of paramount importance to support the hipotesys that SNRs are the
factories of Galactic cosmic rays.
Recent developments in the modeling of the mechanism of diffusive shock
acceleration are discussed, with emphasis on the role of magnetic field
amplification and the presence of neutrals in the circumstellar environment.
Special attention will be devoted to observational consequences of non linear
effects on the multi-wavelength spectrum of SNRs, with emphasis on X-ray and
gamma-ray emission. Finally we also discuss how Balmer lines, detected from
several young SNRs, can be used to estimate the shock dynamical properties and
the efficiency of CR acceleration.
\end{abstract}


\end{frontmatter}

% \linenumbers

\section{Introduction}
\label{sec:intro}
The idea that supernova remnants (SNRs) can generate the bulk of cosmic rays
(CRs) observed at Earth is mainly supported by the fact that from the energetic
point of view SNRs are the only class of sources in the Galaxy that can provide
enough energy to explain the observed flux of CRs. Clearly the energetic
argument alone cannot provide a proof for this {\it SNR paradigm}. The basic
requirements to be fulfilled are: 1) that SNRs should convert a fraction around
10-20\% of their explosion energy into CRs; 2) that the spectrum individual
elements and the consequent all-particle spectrum are well reproduced; 3) that
the chemical abundances of nuclei are well described; 4) that the
multi-frequency observation (from the radio to the gamma-ray band) of individual
SNRs are well described; 5) that the anisotropy observed in the CR arrival
direction is compatible with the distribution of SNRs in the Galaxy.
All these points are subject to current research activities aimed to fit them
in a coherent theoretical framework explaining both particle acceleration and
propagation in the Galaxy.

In this work we underline the main features of the mechanism of particle
acceleration that is thought to work in SNRs (\S~\ref{sec:nldsa}), namely the
diffusive shock acceleration in its non linear version (NLDSA). Than we will
analyze how some observations of SNRs could be used to infer properties on the
acceleration mechanism (\S~\ref{sec:obs}). In particular we will focus on three
different wavebands: 
(\S~\ref{sec:Xray}) detection of thin X-ray filaments;
(\S~\ref{sec:gamma-ray}) observation of gamma-ray spectrum;
(\S~\ref{sec:neutral}) Balmer lines produced by SNR shocks wich expand in
partially ionized medium.

\section{Basic features of NLDSA}
\label{sec:nldsa}
The acceleration mechanism that is usually assumed to work in SNRs is diffusive
shock acceleration (DSA)\cite{bell78,bland-eich87} where particles are
scattered back and forward accross the shock surface by magnetic turbulence,
gaining energy at each cycle. In the so called test-particle limit, where
accelerated particles are assumed to be dynamically unimportant, the theory
predicts a spectrum of accelerated particles wich is a power law in energy
$\propto E^{-\gamma}$ where the index $\gamma= (r+2)/(r-1)$ depends only on the
compression factor, $r$, of the shock and does not depend on the details of the 
particle scattering process. Strong shocks in monoatomic gas has $r=4$, giving
$\gamma=2$.

However the energetic requirement that at least $\sim 10-20\%$ of the
kinetic energy of the supernova shell is converted to CRs leads to realize
immediately that the standard test-particle version of the theory is not
applicable to the description of CR acceleration. The reaction of accelerated
particles onto
the accelerator cannot be neglected and in fact it is responsible for spectral
features that may represent potential signatures of CR acceleration. 

There is another, possibly more important reason why DSA must include the
reaction of accelerated particles: the standard diffusion coefficient typical of
the interstellar medium (ISM) only leads to maximum energies of CRs in the range
of $\sim$GeV, rather than $\sim10^6$GeV (around the knee) required by
observations. A possible way to solve this inconsistency is through the plasma
instabilities induced by the same accelerated particles. In fact these
instabilities can generate the magnetic field structure on which particles may
scatter\cite{lag-ces83}, thereby reducing the acceleration time and reach larger
values of the maximum energy. These plasma instabilities can be effective only
if a non negligible fraction of energy is carried by high energy particles.

The main instability which is thought to be effective is the {\it streaming
instability} induced by CRs which leads to magnetic field amplification where a
spatial gradient of CRs exists. In our case a CR gradient develops upstream of
the shock, hence here the magnetic field is amplified. In the absence of damping
and if only resonant streaming instability is excited, the strength of the
amplified magnetic field can be estimated as a function of the initial field
$B_0$ (by using an extrapolation of quasi-linear theory) as
\begin{equation} \label{eq:Bamp}
 \delta B \approx B_0 \sqrt{2 M_A \xi_{cr}} \,,
\end{equation}
where $M_A$ is the Alfv\`enic Mach number and $\xi_{cr}= P_{cr}/(\rho_0
u_{sh}^2)$ is the pressure of CRs at the shock divided by the total ram
pressure. For typical values of $M_A\sim 1000$ and $\xi_{cr} \simeq 0.20$ we get
$\delta B/B_0\sim 20$ and the field is further amplidied by the compression when
the fluid element crosses the shock from upstream to downstream.

These two processes, dynamical reaction of accelerated particles and magnetic
field amplification, are the two most important ones to take into account when
describing non-linear particle acceleration in SNR shocks.
In the last years many authors have developed a non linear version of DSA
including the back reaction of CRs (see Ref.\cite{mal-drury01} for a review) and
more recently also the self-generation of magnetic field has been included
self-consistently\cite{ama-bla05,caprioli08}.
In the next section we illustrate some important predictions of NLDSA, comparing
them with observations of SNRs in X-ray, gamma-ray and optical H$\alpha$
emission.

\section{Observational evidences of efficient CR acceleration}
\label{sec:obs}

\subsection{Non-thermal X-ray filaments}
\label{sec:Xray}
Non-thermal X-ray radiation is produced in SNRs through synchrotron
emission of high energy electrons in the magnetic field around the shock.
The emission is dominated by the region downstream of the shock where
the magnetic field is stronger and is cut off at a frequency that, in the case
of Bohm diffusion, is independent of the strength of the magnetic field:
\begin{equation}
 \nu_{\max} \approx 0.2 \, u_8^2 \,{\rm keV} \,,
\end{equation}
where $u_8=u_{sh}/(10^8 {\rm cm/s})$ is the shock velocity in units of 1000
km/s. The maximum energy of accelerated electrons depends on the strength of the
amplified magnetic field and can be estimated as
\begin{equation}
 E_{\max} \approx 10 \, B_{100}^{-1/2} u_8 \, {\rm TeV} \,,
\end{equation}
where $B_{100} = B/100\mu$G is the magnetic field in units of 100 $\mu$G.

The emission region has a spatial extent at $\nu \sim \nu_{\max}$ which is
determined by diffusion and can be written as
\begin{equation}
 \Delta x \approx \sqrt{D(E_{\max})\, \tau_{loss}(E_{\max})} 
    \approx 0.04 \, B_{100}^{-3/2} \, {\rm pc} \,.
\end{equation}
The typical thickness of the X-ray rims found through high resolution
observations is of order $\sim 10^{-2}$ pc, thereby leading to predicting values
of the magnetic field of $100-300 \mu$G downstream of the shock.

\begin{figure}
\begin{center}
\resizebox{9cm}{!}{
\includegraphics{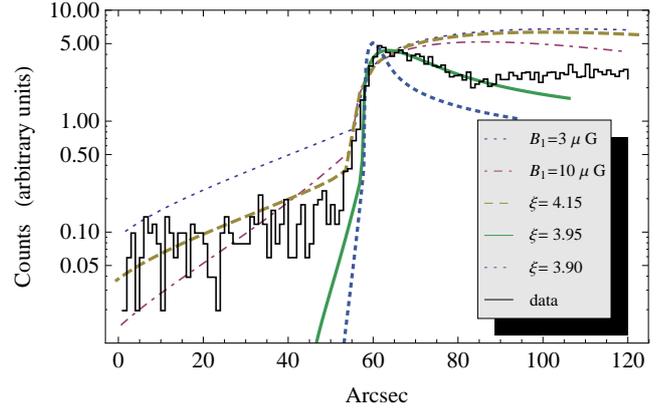}}
\end{center}
\caption{Radial profile of the emission in the 0.8-2.0 keV band, extracted from
a narrow strip around the rim of SN 1006 (thin black line). Overplotted are the
theoretical predictions of NLDSA for 3 different injection efficiencies (thick
lines) and those of test-particle theory for two different values of the
pre-existing turbulent magnetic field upstream (thin lines).}
\label{fig:rim}
\end{figure}

In Fig.~\ref{fig:rim} we show the X-ray brightness profile (histogram) of part
of the rim of SN 1006. The thick lines refer to the predictions of
NLDSA\cite{morlino10} for three values of the injection efficiency (larger
values of $\xi$ correspond to lower efficiencies). The thin lines are the
predicted radial profiles in test-particle theory for two values of the upstream
magnetic field (as indicated). Two pieces of information arise from this figure:
1) the narrow rims ($\sim 10-20$ arcsec) downstream can only be reproduced for
efficient particle acceleration scenarios; 2) the predicted X-ray emission from
the precursor region drops below the background for the same cases in which the
rims are present. In other words, the non-detection of the precursor X-ray
emission might be additional evidence for efficient acceleration. This is due to
the fact that the spatial extent of the precursor is reduced when magnetic
fields are amplified by CR streaming.

Although very interesting, this interpretation is not totally unique:
it could be that the magnetic field is not amplified upstream by CRs, but rather
enhanced due to fluid instabilities downstream in quasi-perpendicular
shocks\cite{giac-jok07}. In this case the non-detection of the upstream
emission could be due to a dominantly perpendicular topology of the magnetic
field lines.

\subsection{$\gamma$-ray observations}
\label{sec:gamma-ray}
A crucial step towards confirming or rejecting the SNR paradigm might
be made through gamma-ray observations both in the TeV energy range,
by using Cherenkov telescopes, and in the GeV energy range accessible to
the Fermi and AGILE gamma-ray telescopes. Gamma radiation can be produced mainly
as a result of inverse Compton scattering (ICS) of relativistic electrons on
the photon background and in inelastic proton-proton scatterings with production
and decay of neutral pions. In the last few years many SNRs have been detected
both with Cherenkov and with gamma-ray telescopes but the present observational
situation is rather puzzling. From one hand the idea that SNRs can accelerate
hadrons has been confirmed, beyond any reasonable doubt, by the detection of a
bump in the GeV spectrum of the remnant W44 by AGILE\cite{giuliani11}.
This bump is in fact interpreted as the threshold of the $\pi^0$ decay,
pointing towards an hadronic origin of the GeV spectrum. On the other
hand the maximum energy of protons inferred from the spectrum of W44 is $\sim$
60-100 GeV, orders of magnitudes below the knee energy of the CR spectrum.

Moreover the gamma-ray spectrum of W44 as well as most spectra observed
by Fermi (see \cite{funk09,tanaka09} for reviews) hint to rather steep spectra
of accelerated particles ($\propto E^{-\gamma}$, with $\gamma \sim 2.3-3$),
which are not easy to accommodate in the context of NLDSA that predicts flat
spectra, possibly even flatter than $E^{-2}$ at high enough
energy\cite{caprioli09}.

The most likely explanation for this discrepancy might lie in a rather subtle
detail of the DSA theory, namely that the velocity relevant for particle
acceleration is the velocity of waves with respect to the plasma. Particles
scatter with magnetic turbulence upstream and downstream af the shock, and feel
an effective compression ratio which is given by
\begin{equation} \label{eq:r_vA}
 r = (u_1 \pm v_{A,1}) / (u_2 \pm v_{A,2}) \,,
\end{equation}
where $u_1$ ($u_2$) and $v_{A,1}$ ($v_{A,2}$) are the plasma and the wave speeds
upstream (downstream), respectively, and the sign depends on the wave
propagation direction. Usually the wave speed is negligible compared with the
plasma velocity in the shock frame, but in the presence of magnetic field
amplification this condition might be weakly violated. This is very bad news in
that the spectral changes induced by this effect depend not only on the wave
speed but on the wave polarization as well (which determines the signs in
Eq.~(\ref{eq:r_vA})).
In \cite{caprioli10,ptu-zir10} the authors show that there are situations in
which the spectral steepening can indeed be sufficient to explain the observed
spectrum of CRs and required by Fermi data on some SNRs.
In fact, assuming that the turbulence is due to Alfv\`en waves induced by the
streaming of CRs, than $v_{A,1}= B_1/\sqrt{4\pi \rho_1}$ is the upstream
Alfv\`en speed, to be taken with negative sign because waves propagate
away from the shock, and $v_{A,2} \approx 0$ because downstream we can expect
efficient helicity mixing. If the magnetic field amplification is strong enough
to produce $v_{A,1}\simeq 0.2 u_{sh}$ than the particle spectral index
becomes $\gamma= (r+2)/(r-1) \simeq 2.3$.

The main caveat in the previous argument is that steep spectra ($\gamma >
2$) imply low energy in the highest energy particles. But those particles are
the only ones that, through the streaming instability, can produce turbulence
with the right wavelength on which they scatter efficiently. If the energy
density in those waves is not large enough the scattering is not efficient and
particles can escape the system without reaching the required energy to explain
the CR spectrum. The only way to test this mechanism is to develop consistent
model of particle-wave interaction in the contest of NLDSA and compare the
predictions with observations.

\begin{figure}
\begin{center}
\resizebox{9cm}{!}{
\includegraphics{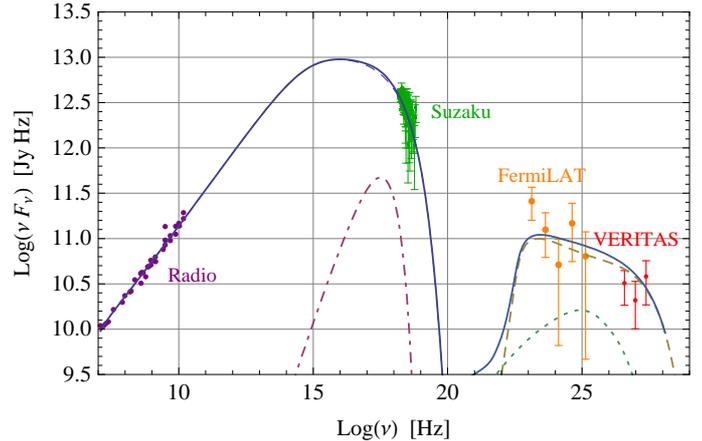}}
\end{center}
\caption{Spatially integrated spectral energy distribution of Tycho's SNR. The
curves show synchrotron emission (thin dashed), thermal electron bremsstrahlung
(dot-dashed), pion decay (thick dashed) and ICS (dotted) as calculated using
NLDSA. The total sum is showed by the solid curve. The experimental data from
Fermi-LAT and VERITAS include only statistical error at 1 $\sigma$. 
The figure is from {\protect\cite{tycho}}.}
\label{fig:tycho}
\end{figure}

Indeed this idea has been successfully applied to the Tycho's
remnant\cite{tycho} which is, up to date, the only convincing SNR showing
evidence for efficient CR acceleration from the gamma-ray observation. Tycho is
a young type-I/a SNR detected in many wavebands, from radio up to TeV.
In \cite{tycho} the NLDSA has been coupled to magnetic field amplification
(including the wave speed effect) and to the radiative processes in a self
consistent way, in order to explain the observed multi-wavelength spectrum.
Fig.~(\ref{fig:tycho}) shows the result. All data are well explained by the
model. The steepening induced by the wave speed with respect to the plasma
results in a gamma-ray spectrum $\propto E^{-2.2}$ which well account for the
GeV-TeV data. At the same time the streaming instability plus the shock
compression results in a downstream magnetic field $\sim 300\mu$G which well
account for the integrated radio and X-ray emission as well as for their
morphology. Remarkably the maximum energy inferred for accelerated protons is
$\sim 500$ TeV, only a factor 6 below the knee energy, while the total energy
converted into accelerated particles is $\sim 12\%$ of the bulk kinetic energy
of the expanding shock, which is, again, what is required in order to explain
the CR flux observed at Earth.

\subsection{Acceleration in partially neutral plasmas and Blamer emission}
\label{sec:neutral}
Shocks produced by supernaova explosions are {\it collisionless}, which means
that the fluid discontinuities are formed by electromagnetic instabilities while
particle collisions are negligible because the low density of interstellar
medium. On the other hand the interstellar medium itself is often only partially
ionized. As we will explain in a while, the presence of neutrals can
significantly modify the shock structure in a way that up to know has been
completely neglected in the shock theory. As a consequence, the spectrum of
accelerated particles can also be affected.

The neutral component interacts with the ionized component only through the
processes of charge exchange and ionization.  These two processes lead to
exchange of energy and momentum between charged and neutral particles both
upstream and downstream of the shock. In particular, neutrals that suffer a
charge exchange downstream with shock-heated ions generate high velocity
neutrals that have a finite probability of returning upstream.
These neutrals might then deposit heat in the upstream plasma through ionization
and charge exchange, thereby reducing the fluid Mach number. This phenomenon,
leads to a reduction of the shock compression factor and to the formation of a
{\it neutral-induced} precursor upstream of the shock \cite[]{neutrals}. The
scale length of the precursor is determined by the ionization and charge
exchange interaction lengths of fast neutrals moving towards upstream infinity.
In the case of a shock propagating in the interstellar medium, the effects of
ion-neutral interactions are especially important for shock velocities $< 3000$
km/s. What is remarkable is that, when we consider the acceleration of
particles, the reduction of the sub-shock compression factor (induced by
neutrals) to values $<4$ can result in a spectrum of accelerated particles
steeper than $E^{-2}$.
This effect is energy dependent and is more effective for particles whose
diffusion length ahead of the shock is $\lesssim$ than the ionization or charge
exchange interaction lengths.

\begin{figure}
\begin{center}
\resizebox{9cm}{!}{
\includegraphics{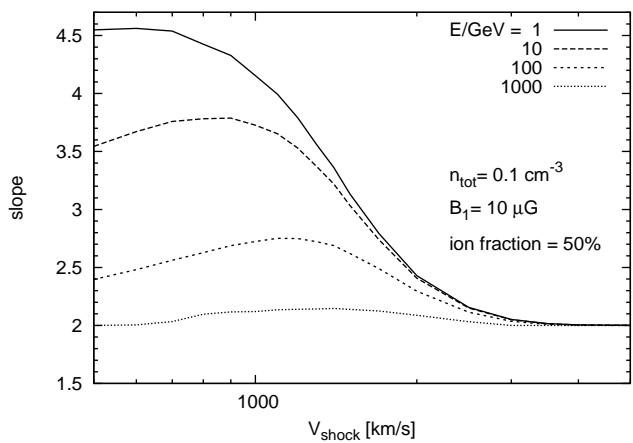}}
\end{center}
\caption{Slope of the spectrum of accelerated particles in case of shock
propagating into a plasma with 50\% ionization fraction and total density of 1
particle per cm$^{-3}$. Calculations are performed in test particle limit for
energies $E = 1$, 100, 100, 1000 GeV, and as a function of the shock speed. The
presence of neutrals produces a deviation from the standard slope equal to $2$.}
\label{fig:slope}
\end{figure}

The structure of shocks in presence of neutrals can be described using the
kinetic theory where the coupling of neutral particles with ions and electrons
through charge-exchange and ionization is described by the the Vlasov
equation.
In Fig.~(\ref{fig:slope}) we report a preliminary result of this effects: we
computed the slope of accelerated particles in the test-paricle case considering
a total density of 1 particle per cm$^{3}$ with 50\% of ionization fraction
and an upstream magnetic field of $10\mu$G. We see that for shock speed $< 3000$
km/s even particles at 1 TeV can have a slope steeper than 2.

In principle the presence of neutrals can provide an explanation for the steep
spectra observed in GeV band from several SNRs (see \S\ref{sec:gamma-ray}).
Nevertheless these results should be taken with care because are computed in
test particle theory; when the effect of CR will be included, non linear
effects could change the conclusion even dramatically.

Apart of the effect that neutrals have on the shock dynamics, there is a second
reason why they are worth to study. Hydrogen atoms that undergo charge exchange
or collisional excitation produce Balmer and Lyman lines. These lines have been
detected from several SNR shocks (especially Balmer) which, for this reason, are
called {\it Balmer-dominated} shocks. What is remarkable is that Balmer emission
can be used to infer some shock properties, including the acceleration
efficiency, as we will explain in the following.

The emission lines typically have two components. The first one is a narrow line
whose width is characteristic of the cold interstellar medium and results from
direct excitation of neutral hydrogen atoms. The second component has a much
broader line width, and arises from a second population of hydrogen atoms
created by charge exchange process between cold pre-shock hydrogen and hot
shocked protons. These hot atoms can be produced in an excited state or can be
excited by subsequent collisions with protons or electrons. Hence the line width
of the broad component is characteristic of the thermal velocity of the shocked
protons.

The situation becomes more interesting in the presence of CRs:
when the pressure of accelerated particles upstream of the shock generates
the precursor, in the shock frame the bulk motion of ions is slowed down
with respect to neutrals. Neutral atoms keep moving with the velocity that
ions have at upstream infinity, $u_{sh}$, therefore a difference in bulk
velocity between the two species arises. At the shock crossing, the ionized
plasma suffers shock heating and compression while the neutral component does
not feel the shock and keeps moving with speed $u_{sh}$ (with respect ot the
shock frame). In the presence of CR acceleration, the temperature of the ions
downstream is clearly lower that in the absence of CRs, simply because of energy
conservation (there is now another component, CRs, into which the ram pressure
$\rho_0 u_{sh}^2$ can be channeled).
Therefore one can expect that the width of the broad
Balmer line is somewhat smaller than in the absence of CRs. This phenomenon has
been recently observed in the SNR RCW86\cite{helder11}.

Measurements of the proper motion of the shock lead to a shock velocity
of $u_{sh} = 6000 \pm 2800$ km/s, which in turn should imply a downstream
temperature of $T_2 = 20-150$ keV or $T_2 = 12-90$ keV depending on whether
protons and electrons reach or not thermal equilibrium downstream, and
assuming that no CR acceleration is taking place. The temperature inferred
from the width of the broad Balmer line is $T_2 = 19.2\pm1.1$ keV. The authors
interpreted this discrepancy as the result of effective CR acceleration at
the SNR shock, with an estimated efficiency of $\gtrsim 15\%$ considering an
electron temperature $<70\%$ of the proton temperature.

In CR modified shocks also the width of the narrow line should be modified. In
fact some level of charge exchange is expected to take place in the precursor
because of the difference in bulk velocity between the ionized plasma and the
neutrals. The process is responsible for producing a population of warmer
neutrals upstream. After shock crossing this leads to a narrow Balmer line which
is broader than in the absence of a CR induced precursor. This effect was
observed in several SNRs\cite{sollerman03}. An independent signature of the same
phenomenon was recently found in the Tycho's SNR\cite{lee10}. The authors used
the Hubble Space Telescope to measure the intensity of the broadened narrow
Balmer line as a function of position around the shock and claim that there is
substantial emission from the region in front of the shock. This finding could
be either the signature of different bulk velocities between ions and neutrals
(precursor) or of a different temperature of the two components. In this latter
case, the ions could have been heated up due to the action of turbulent heating
in the precursor.

\section{Conclusions}
We provided a short review of the particle acceleration theory around SNR
shocks underlining the fact that accelerated particles cannot be passive
spectators in the acceleration process, but they should have an important
dynamical role. When the back-reaction of those particle is taken into account,
the theory makes several predictions which seems to be confirmed by recent
observational findings.
Probably the most streaking one is the detection of narrow X-ray filaments
close to the shocks, which are interpreted as due to rapid electron synchrotron
losses, supporting the idea that the magnetic field is strongly amplified as
predicted by NLDSA.
 
On the other hand gamma-ray observations challenges the current theory because
the observed spectra in the GeV-TeV range are typically steeper than what NLDSA
predicts. Nevertheless we showed that this discrepancy can be accounted for at
least in two different ways. One possibility is to take into account the finite
velocity of waves responsible for particle scattering which, under certain
circumstances, reduces the effective compression ratio felt by particles,
resulting in steeper spectra.
The second possibility involves the role of neutral particles which has never
been properly taken into account in shock theory. We showed that neutrals
can modify the shock structure through the processes of charge exchange
and ionization with the ionized component, producing a neutral-induced
precursor ahead of the shock. Also in this case the compression ratio can be
remarkably reduced and steeper spectra are produced.
Both these ideas still require to be tested in a full self-consistent theory.

The inclusion of neutrals in the shock theory has a second important
consequence. Hydrogen atoms that undergo excitation and charge-exchange
in the shock region produce Balmer and other optical lines. These lines can be
used to infer the temperature of the shocked gas and also to reveal the
presence of a precursor providing an estimate of the CR acceleration efficiency.

% \bibliographystyle{elsarticle-num}
% \bibliography{<your-bib-database>}

\end{document}